# Analysis of the September ε-Perseid outburst in 2013

**José M. Madiedo[1], Jaime Zamorano[2], Josep M. Trigo-Rodríguez[3, 4], José L. Ortiz[5], José A. Docobo[6], Jaime Izquierdo[2], Juan Lacruz[7], Pedro P. Campo[6], Manuel Andrade[8, 6], Sensi Pastor[9], José A. de los Reyes[9], Francisco Ocaña[2, 10], Alejandro Sánchez-de Miguel[11, 2], Pep Pujols[12]**

[1] Facultad de Ciencias Experimentales, Universidad de Huelva. 21071 Huelva, Spain.

[2] Dpto. de Física de la Tierra y Astrofísica, Facultad de Ciencias Físicas, Universidad Complutense de Madrid, 28040 Madrid, Spain.

[3] Institute of Space Sciences (CSIC), Campus UAB, Facultat de Ciències, Torre C5-parell-2ª, 08193 Bellaterra, Barcelona, Spain.

[4] Institut d'Estudis Espacials de Catalunya (IEEC), Edif.. Nexus, c/Gran Capità, 2-4, 08034 Barcelona, Spain

[5] Instituto de Astrofísica de Andalucía, CSIC, Apt. 3004, Camino Bajo de Huetor 50, 18080 Granada, Spain.

[6] Observatorio Astronómico Ramón María Aller (OARMA). Universidade de Santiago de Compostela, Avenida das Ciencias, Campus Vida. Santiago de Compostela, Spain.

[7] La Cañada Observatory (MPC J87), Ávila, Spain.

[8] Departamento de Matemática Aplicada. Escola Politécnica Superior de Enxeñaría, Universidade de Santiago de Compostela, Campus Universitario, 27002 Lugo, Spain.

[9] Observatorio Astronómico de La Murta. Molina de Segura, 30500 Murcia, Spain.

[10] Quasar Science Resources, S. L., Las Rozas de Madrid. 28232 Madrid, Spain.

[11] Environment and Sustainability Institute, University of Exeter, Penryn, Cornwall TR10 9FE, U.K.

[12] Agrupació Astronómica d'Osona (AAO), Carrer Pare Xifré 3, 3er. 1a. 08500 Vic, Barcelona, Spain

## ABSTRACT

We analyze the outburst experienced by the September ε-Perseid meteor shower on 9 September 2013. As a result of our monitoring the atmospheric trajectory of 60 multi-station events observed over Spain was obtained and accurate orbital data were derived from them. On the basis of these orbits, we have tried to determine the likely parent body of this meteoroid stream by employing orbital dissimilarity criteria. In addition, the emission spectra produced by two events belonging to this meteor shower were also recorded. The analysis of these spectra has provided information about the chemical nature of their progenitor meteoroids. We also present an estimation of the tensile strength for these particles.

## 1 INTRODUCTION

The September ε-Perseid (SPE) meteoroid stream gives rise to an annual display of meteors from about September 7 to September 23, peaking around September 12 (Jenniskens 2006). This minor shower was first observed by Denning (1882), and is currently included in the IAU list of meteor showers with code 208 SPE. No systematic analysis of this shower was performed during the early to mid twentieth century, and the first reliable data about this stream were analyzed in Hoffmeister (1948). The next observations were published by Trigo-Rodríguez (1989), who clearly identified SPE activity over the sporadic background, with often trained and bright meteors exhibiting a peak zenithal hourly rate ZHR = 5 meteors $h^{-1}$ in 1989.

Only two outbursts of SPE meteor activity have been reported. The first of these was unexpected and took place on 9 September 2008, with an activity consisting mostly of bright meteors (Jenniskens et al. 2008; Rendtel and Molau 2010). This outburst was not favourable for observers in Europe. So, despite our systems were monitoring the night sky, we could not record this activity increase. The second SPE outburst occurred on 9 September 2013. It took place between 21h30m and 23h20m UT and was confirmed in Jenniskens (2013). On the basis of the results obtained from the analysis of the 2008 outburst, and by assuming that SPE meteoroids were produced by a long-period comet that ejected these particles before the year 1800 AD, Jenniskens (2013) inferred that this dust trail should encounter Earth on 9 September 2013 at 22h15m UT. This is in good agreement with the circumstances of the 2013 SPE outburst. However, the parent comet of this stream has not been identified yet. Accurate orbital data obtained from the analysis of SPE meteors could help to find the likely parent of the September ε-Perseids. And meteor spectroscopy can also play an important role to derive information about the chemical nature of these meteoroids and their progenitor body.

Optimal weather conditions over most of the Iberian Peninsula during the first half of September 2013 allowed us to analyze the meteor activity produced by the SPE stream. In this work we focus on the analysis of the 2013 SPE outburst. From our recordings we have obtained orbital information about meteoroids belonging to this poorly known stream. The tensile strength of these particles is also estimated. Besides, two emission spectra produced by SPE meteors are also analyzed. These are, to our knowledge, the first SPE spectra discussed in the scientific literature.

## 2 INSTRUMENTATION AND DATA REDUCTION TECHNIQUES

The meteor observing stations that were involved in the monitoring of the September ε-Perseid outburst analyzed here are listed in Table 1. These employ between 3 and 12 high-sensitivity CCD video cameras (models 902H and 902H Ultimate from Watec Co., Japan) to monitor the night sky (Madiedo & Trigo-Rodríguez 2008; Trigo-Rodríguez et al. 2009). Their field of view ranges from 90 x 72 degrees to 14 x 11 degrees. These CCD devices work according to the PAL video standard and, so, they generate interlaced video imagery at 25 fps with a resolution of 720x576 pixels. More details about these devices and the way they are operated are given, for instance, in (Madiedo 2014). In order to obtain the atmospheric trajectory of the meteors and the heliocentric orbit of the progenitor meteoroids we have employed the AMALTHEA software (Madiedo et al. 2013a,b), which follows the methods described in Ceplecha (1987).

To record meteor emission we have attached holographic diffraction gratings (with 500 or 1000 lines/mm, depending on the device) to the objective lens of some of the above-mentioned CCD video cameras. With these slitless videospectrographs we can record the emission spectrum of meteors brighter than magnitude -3/-4 (Madiedo et al. 2013c; Madiedo 2014). The analysis of the emission spectra obtained during the SPE observing campaign analyzed here was performed by means of the CHIMET software (Madiedo et al. 2013c).

## 3 OBSERVATIONS AND RESULTS

In 2013 our meteor observing stations observed activity from the September ε-Perseids from September 1 to September 12. On 9 September 2013, at about 21h35m UT, our CCD video devices registered a marked increase in meteor activity associated with this stream, including some fireballs. A careful checking of these data confirmed the SPE outburst between around 21h35m UT on September 9 and 0h 20m UT on September 10, in good agreement with the circumstances described in Jenniskens (2013). From the analysis of the multi-station events recorded from sites listed in Table 1 we have obtained the atmospheric trajectory of these meteors. However, we just took into consideration those trails for which the convergence angle was above 20 degrees. This parameter, which is usually employed to measure the quality of the results, is the angle between the two planes delimited by the observing sites and the meteor atmospheric path (Ceplecha 1987). A total of 60 SPE meteors satisfied this condition. These events are listed in Table 2, which shows the absolute peak magnitude (M), the initial (preatmospheric) photometric mass of the progenitor meteoroid ($m_p$), the initial ($H_b$) and final ($H_e$) heights of the meteor, the right ascension ($\alpha_g$) and declination ($\delta_g$) of the geocentric radiant (J2000.0), and the preatmospheric ($V_\infty$) and geocentric ($V_g$) velocities. To identify each meteor, we have employed a code with the format DDYYEE, where DD is the day of the month (which ranges between 01 and 12 for the meteors analyzed here), and YY corresponds to the last two digits of the recording year. The two digits EE are employed to number meteors recorded during the same night, so that 00 is assigned to the first meteor imaged, 01 to the second one and so on. The averaged value for the observed initial (preatmospheric) velocity was $V_\infty = 65.9 \pm 0.2$ km $s^{-1}$. The photometric mass of the parent meteoroids ranged between 0.01 to 16 g (Table 2). The orbital parameters derived for the meteoroids that gave rise to these meteor events are listed in Table 3.

## 4 DISCUSSION

### 4.1 Parent body

The averaged orbital data calculated by taking into account a total of N = 60 SPE orbits are shown in Table 3. This table also includes the average orbit of meteors observed during the outburst (N=28 meteors). As can be noticed, the difference between both averaged orbits is not significant. With these parameters we have obtained that the value of the Tisserand parameter with respect to Jupiter yields $T_J = -0.65 \pm 0.44$. This agrees with the assumption in Jenniskens (2013) that SPE meteoroids are produced by a long period comet.

Besides, we have calculated the so-called $K_B$ parameter, which according to Ceplecha (1988) can be employed to classify meteoroids into four different populations: A-group, comprising particles similar to carbonaceous chondrites ($7.3 \leq K_B < 8$); B-group of dense cometary material ($7.1 \leq K_B < 7.3$); C group of regular cometary material ($6.6 \leq K_B < 7.1$); and D-group of soft cometary material ($K_B < 6.6$). This parameter is defined by the following equation:

$$K_B = \log \rho_B + 2.5 \log V_\infty - 0.5 \log \cos z_R + 0.15 \quad (1)$$

where $\rho_B$ is the air density at the beginning of the luminous trajectory (in g cm$^{-3}$), $V_\infty$ is the preatmospheric velocity of the meteoroid (in cm s$^{-1}$), and $z_R$ is the inclination of the atmospheric trajectory with respect to the vertical. We have obtained the air density $\rho_B$ by using the NRLMSISE-00 atmosphere model (Picone et al. 2002). According to our computations, the average $K_B$ parameter for the SPE events in Table 2 yields $K_B = 6.9 \pm 0.2$. This result suggests that meteoroids in this stream belong to the group of regular cometary materials.

We have tried to determine the likely parent comet of SPE meteoroids by means of orbital dissimilarity criteria (Williams 2011). In this approach we have employed the ORAS program (ORbital Association Software) to search through the Minor Planet Center database in order to establish a potential link between the SPE stream and other bodies in the Solar System (Madiedo et al. 2013d). This analysis has been performed by calculating the Southworth and Hawkings $D_{SH}$ criterion (Southworth & Hawkins 1963). However the lowest values obtained for the $D_{SH}$ function are of about 1.50, which is well above the $D_{SH}$ <0.15 cutoff value usually adopted to validate a potential association (Linblad 1971a,b). So we conclude that the parent comet of the SPE stream is not catalogued.

## 4.2. Meteor initial and final heights

The dependence with meteoroid mass of the beginning and final heights of meteors analyzed in this work has been plotted in Figures 1 and 2, respectively. As can be seen in Figure 1, the initial height $H_b$ increases with increasing meteoroid mass. This behaviour has been also found for other meteor showers with a cometary origin (see, e.g., Koten et al. 2004, Jenniskens 2004, Madiedo 2015). We have described this behaviour by means of a linear relationship between $H_b$ and the logarithm of the meteoroid photometric mass (solid line in Figure 1). The slope of this line is 2.38 ± 0.70. According to this result, the increase of the beginning height with meteoroid mass is less important for the September ε-Perseids than for the Leonids (a = 9.9 ± 1.5), the Perseids (a = 7.9 ± 1.3), the Taurids (a = 6.6 ± 2.2), and the Orionids (a = 5.02 ± 0.65) (Koten et al. 2004). But more pronounced than for the ρ-Geminids (a = 1.1 ± 0.5), which are produced by tough cometary meteoroids (Madiedo 2015), and the Geminids (a = 0.46 ± 0.26) (Koten et al. 2004), which have an asteroidal origin (Jenniskens 2004). Despite the preatmospheric velocity of the September ε-Perseids and the Perseids is similar (~65 km $s^{-1}$ and ~61 km $s^{-1}$, respectively), the beginning height exhibited by SPE meteors is significantly lower. Thus, for instance, for SPE meteors $H_b$ is of below 110 km for a meteoroid mass of about 0.02 g (Figure 2), but ~120 km for Perseid members with the same mass (Koten et al. 2004). This suggests that SPE meteoroids are tougher.

The larger is the meteoroid mass, the lower is the terminal point $H_e$ of the meteor (Figure 2). The slope of the line we have employed to model this behaviour (solid line in Figure 2) yields -2.25 ± 0.45. Since slower meteoroids tend to penetrate deeper in the atmosphere, it is not surprising that SPE meteoroids, which exhibit an initial velocity of ~65 km $s^{-1}$, do not penetrate as deep as the Perseids with a preatmospheric velocity of ~61 km $s^{-1}$ (Koten 2004), the Geminids with a velocity of ~36 km $s^{-1}$ (Jenniskens 2004), or the ρ-Geminids with a velocity of about 23 km $s^{-1}$ (Madiedo 2015).

## 4.3 Meteoroid strength

The tensile strength of meteoroids ablating in the atmosphere can be estimated by analyzing the flares exhibited by the corresponding meteors. According to this approach, these flares take place as a consequence of the sudden break-up of the meteoroid when the aerodynamic pressure overcomes the strength of the particle (Trigo-Rodríguez & Llorca 2006). However, SPE events listed in Table 2 exhibited a quasi-continuous ablation behaviour, with smooth lightcurves that revealed that no flares occurred during their interaction with the atmosphere. So, we have employed this technique to evaluate the maximum aerodynamic pressure suffered by SPE meteoroids, which has provided a lower limit for their tensile strength. This aerodynamic pressure S can be estimated by using the following relationship (Bronshten 1981)

$$S=\rho_{atm}\cdot v^2 \qquad (2)$$

where $\rho_{atm}$ and v are the atmospheric density and meteor velocity at a given height, respectively. In this work we have calculated the air density by employing the NRLMSISE-00 atmosphere model (Picone et al. 2002). From the analysis of the atmospheric trajectory calculated for meteors in Table 2 we have obtained a maximum aerodynamic pressure of $(2.9 \pm 0.3) \cdot 10^5$ dyn $cm^{-2}$. This value is higher than the average strength found for Quadrantid and Perseid meteoroids ($\sim 2 \cdot 10^5$ dyn $cm^{-2}$ and $(1.2 \pm 0.3) \cdot 10^5$ dyn $cm^{-2}$, respectively), and below the strength of the Taurids ($(3.4 \pm 0.7) \cdot 10^5$ dyn $cm^{-2}$) (Trigo-Rodríguez & Llorca 2006, 2007).

## 4.4 Emission spectra

The video spectrographs operated at stations 1 to 6 in Table 1 recorded a total of 8 SPE emission spectra during the outburst recorded on September 9-10. Unfortunately 6 of these were too dim to be analyzed, but the other two had enough quality to provide information about the chemical nature of these meteoroids. These spectra were produced by meteors labelled as SPE091327 and SPE091331 in Table 2, respectively. They have been analyzed with the CHIMET software (Madiedo et al. 2013c), which first deinterlaces the video files containing these signals. Then, the software performs a dark-frame subtraction, and each video frame is flat-fielded. Next, the calibration in wavelength is achieved by identifying emission lines typically found in meteor spectra. Then, the intensity of the signals is corrected by taking into account the spectral efficiency of the spectrograph. The results obtained from this procedure are shown in Figures 3 and 4. In these plots, the most remarkable multiplets have been labelled according to the notation given by Moore (1945). The most noticeable emissions correspond to the atmospheric O I line at 777.4 nm, and to the K and H lines of Ca II-1 at 393.3 and 396.8 nm, respectively. These two lines produced by ionized calcium appear blended in the spectrum. Other prominent contributions are those of Fe I-41 (440.4 nm), the Mg I-2 triplet (517.3 nm), and the Na I-1 doublet (588.9 nm). The emission of atmospheric $N_2$ bands was also identified in the red region of the spectrum.

As in previous works (see e.g. Madiedo et al. 2014), we have investigated the chemical nature of the progenitor meteoroids by analyzing the relative intensity of the Na I-1, Mg I-2 and Fe I-15 multiplets in these spectra (Borovička et al. 2005). To perform this analysis, the intensity (in arbitraty units) of the emission lines associated with these multiplets was measured frame by frame and subsequently corrected according to the efficiency of the spectrograph. Next the contributions in each video frame were added to obtain the integrated intensity for each emission line along the meteor path. For the SPE091327 spectrum the integrated intensities of the Na I-1, Mg I-2 and Fe I-15 multiplets yield 15, 23 and 7, respectively. For the SPE091331 spectrum these intensities yield 35, 55 and 21. In this way we have obtained for the SPE091327 and SPE091331 spectra a Na to Mg intensity ratio of 0.62 and 0.63, respectively. And the Fe/Mg intensity ratio yields 0.29 and 0.38, respectively. The ternary diagram in Figure 5 shows the relative intensity of the emission from the Na I-1, Mg I-2 and Fe I-15 multiplets for both spectra. The solid curve in this plot corresponds to the expected relative intensity, as a function of meteor velocity, for chondritic meteoroids (Borovička et al. (2005)). The position on this solid line corresponding to the velocity of SPE meteors (~65 km $s^{-1}$) is not explicitly specified in the work published by Borovička et al. (2005) (see Figure 6 in that work), although the authors of that paper indicate that the points describing these high-speed meteors are located near the left edge of this curve. By taking this into account we conclude that the points in this diagram describing both spectra show that SPE meteoroids can be considered as normal according to the classification given by Borovička et al. (2005). Thus, the position of these experimental points fit fairly well the expected relative intensity for chondritic meteoroids for a meteor velocity of ~65 km $s^{-1}$.

## 5 CONCLUSIONS

We have analyzed the meteor activity associated with the September ε-Perseid meteoroid stream in 2013. In this context we have observed the outburst experienced by this meteor shower on September 9-10. From the analysis of our recordings we have reached the following conclusions:

1) The dependence with meteoroid mass of the initial height observed for SPE meteors reveals a cometary origin for this stream. The increase of the beginning height with mass is less important for the September ε-Perseids than for the Leonids, the Perseids, the Taurids, and the Orionids. But more pronounced than for the ρ-Geminids and the Geminids The analysis of the $K_B$ parameter suggests that SPE meteoroids consist of regular cometary material.
2) The orbital data calculated from the analysis of our double-station meteors support the idea that SPE meteoroids are associated with a long period comet. However, no parent body could be identified among the objects currently included in the Minor Planet Center database. From this we conclude that the progenitor comet of this meteoroid stream is not yet catalogued.
3) The tensile strength of these meteoroids has been constrained. According to our calculations, the maximum aerodynamic pressure suffered by SPE meteoroids is higher than the tensile strength found for Quadrantid and Perseid meteoroids.
4) We have recorded 8 emission spectra produced by SPE meteors during the outburst recorded on September 9-10. Two of these had enough quality to be analyzed, and these suggest a chondritic nature for SPE meteoroids.

## ACKNOWLEDGEMENTS

We acknowledge support from the Spanish Ministry of Science and Innovation (projects AYA2015-68646-P and AYA 2015-67175-P).

# TABLES

Table 1. Geographical coordinates of the SPMN meteor observing stations that recorded the 2013 outburst of the SPE meteor shower.

| Station # | Station name | Longitude | Latitude (N) | Altitude (m) |
|---|---|---|---|---|
| 1 | Sevilla | 5º 58' 50" W | 37º 20' 46" | 28 |
| 2 | La Hita | 3º 11' 00" W | 39º 34' 06" | 674 |
| 3 | Huelva | 6º 56' 11" W | 37º 15' 10" | 25 |
| 4 | Sierra Nevada | 3º 23' 05" W | 37º 03' 51" | 2896 |
| 5 | El Arenosillo | 6º 43' 58" W | 37º 06' 16" | 40 |
| 6 | Cerro Negro | 6º 19' 35" W | 37º 40' 19" | 470 |
| 7 | Ávila | 4º 29' 30" W | 40º 36' 18" | 1400 |
| 8 | Villaverde del Ducado | 2º 29' 29" W | 41º 00' 04" | 1100 |
| 9 | Madrid-UCM | 3º 43' 34" W | 40º 27' 03" | 640 |
| 10 | La Murta | 1º 12' 10" W | 37º 50' 25" | 400 |
| 11 | Folgueroles | 2º 19' 33" E | 41º 56' 31" | 580 |
| 12 | Montseny | 2º 32' 01" E | 41º 43' 47" | 194 |
| 13 | Lugo | 7º 32' 41" W | 42º 59' 35" | 418 |
| 14 | OARMA | 8º 33' 34" W | 42º 52' 31" | 240 |

Table 2. Trajectory and radiant data (J2000) for the double-station September ε-Perseid meteors discussed in the text.

| Meteor code | Date | Time (UT) ±0.1s | M ±0.5 | $m_p$ (g) | $H_b$ (km) ±0.5 | $H_e$ (km) ±0.5 | $\alpha_g$ (º) | $\delta_g$ (º) | $V_\infty$ (km s$^{-1}$) | $V_g$ (km s$^{-1}$) |
|---|---|---|---|---|---|---|---|---|---|---|
| 011301 | Sep 1 | 4h16m25.8s | -2.0 | 0.17 ± 0.07 | 109.4 | 88.0 | 41.31±0.10 | 38.15±0.02 | 66.2±0.2 | 65.2±0.2 |
| 011302 | Sep 1 | 20h50m52.3s | 1.2 | 0.01± 0.01 | 109.5 | 100.6 | 42.45±0.07 | 37.04±0.02 | 66.9±0.2 | 65.7±0.2 |
| 011303 | Sep 1 | 23h22m20.4s | -2.1 | 0.19 ± 0.08 | 116.6 | 99.5 | 41.88±0.08 | 37.45±0.02 | 66.6±0.2 | 65.4±0.2 |
| 021301 | Sep 2 | 1h15m45.0s | -0.5 | 0.04± 0.01 | 111.2 | 90.5 | 42.03±0.10 | 38.14±0.04 | 66.5±0.2 | 65.3±0.2 |
| 021302 | Sep 2 | 2h44m43.2s | -1.2 | 0.08 ± 0.03 | 112.1 | 95.6 | 42.08±0.09 | 37.72±0.06 | 66.2±0.2 | 65.1±0.2 |
| 031301 | Sep 3 | 1h25m30.3s | -2.7 | 0.35 ± 0.14 | 113.3 | 92.0 | 41.58±0.12 | 36.55±0.10 | 66.3±0.2 | 65.1±0.2 |
| 031302 | Sep 3 | 3h53m12.9s | -0.7 | 0.05 ± 0.02 | 109.9 | 91.4 | 42.13±0.15 | 37.86±0.09 | 66.0±0.2 | 65.0±0.2 |
| 041301 | Sep 4 | 1h04m43.8s | 0.5 | 0.01 ± 0.01 | 109.5 | 98.8 | 43.11±0.14 | 36.74±0.10 | 66.4±0.2 | 65.2±0.2 |
| 051301 | Sep 5 | 1h17m19.2s | -2.5 | 0.29 ± 0.12 | 113.9 | 87.6 | 44.09±0.15 | 37.62±0.10 | 66.5±0.2 | 65.3±0.2 |
| 051302 | Sep 5 | 21h11m18.1s | -3.6 | 0.86 ± 0.35 | 115.6 | 101.8 | 44.24±0.14 | 38.23±0.10 | 66.1±0.2 | 64.8±0.2 |
| 061301 | Sep 6 | 4h59m48.6s | -3.9 | 1.16 ± 0.47 | 114.1 | 90.1 | 44.48±0.10 | 38.10±0.10 | 65.8±0.2 | 64.9±0.2 |
| 081301 | Sep 8 | 2h21m15.4s | 1.1 | 0.01 ± 0.01 | 107.6 | 96.9 | 45.39±0.07 | 38.72±0.08 | 65.7±0.2 | 64.6±0.2 |
| 081302 | Sep 8 | 3h29m08.9s | -3.6 | 0.86 ± 0.35 | 110.4 | 89.8 | 46.47±0.12 | 38.40±0.09 | 65.9±0.2 | 64.9±0.2 |
| 081303 | Sep 8 | 4h00m40.7s | -3.4 | 0.71 ± 0.29 | 115.4 | 84.4 | 45.97±0.15 | 38.26±0.12 | 65.8±0.2 | 64.8±0.2 |
| 091301 | Sep 9 | 0h13m42.3s | 0.8 | 0.01 ± 0.01 | 105.5 | 89.0 | 47.28±0.09 | 38.99±0.10 | 66.1±0.2 | 64.9±0.2 |
| 091302 | Sep 9 | 0h52m35.4s | 1.0 | 0.01 ± 0.01 | 111.0 | 95.4 | 47.16±0.09 | 39.01±0.08 | 66.0±0.2 | 64.8±0.2 |
| 091303 | Sep 9 | 1h31m29.7s | -2.1 | 0.20 ± 0.08 | 113.3 | 97.6 | 47.20±0.10 | 38.94±0.09 | 66.1±0.2 | 65.0±0.2 |
| 091304 | Sep 9 | 1h37m55.5s | 1.5 | 0.01 ± 0.01 | 101.1 | 96.8 | 47.12±0.08 | 39.00±0.08 | 65.9±0.2 | 64.7±0.2 |
| 091305 | Sep 9 | 2h34m02.9s | -0.9 | 0.06 ± 0.02 | 109.9 | 95.4 | 47.54±0.10 | 38.92±0.09 | 66.0±0.2 | 64.9±0.2 |
| 091306 | Sep 9 | 21h43m34.0s | -1.2 | 0.08 ± 0.03 | 107.6 | 97.2 | 47.78±0.09 | 39.28±0.09 | 66.0±0.2 | 64.8±0.2 |
| 091307 | Sep 9 | 21h53m24.2s | -5.1 | 3.87 ± 1.57 | 116.1 | 94.8 | 47.76±0.10 | 39.53±0.10 | 65.9±0.2 | 64.6±0.2 |
| 091308 | Sep 9 | 21h56m32.7s | -1.9 | 0.16 ± 0.06 | 116.6 | 105.0 | 47.84±0.10 | 39.17±0.09 | 66.1±0.2 | 64.8±0.2 |
| 091309 | Sep 9 | 21h56m51.3s | -3.4 | 0.72 ± 0.29 | 114.3 | 85.1 | 47.92±0.08 | 39.76±0.06 | 65.9±0.2 | 64.7±0.2 |
| 091310 | Sep 9 | 21h58m13.5s | 0.1 | 0.01 ± 0.01 | 101.7 | 97.1 | 47.72±0.09 | 39.75±0.09 | 65.9±0.2 | 64.6±0.2 |
| 091311 | Sep 9 | 22h00m44.5s | 0.3 | 0.02± 0.01 | 101.5 | 96.9 | 47.51±0.09 | 39.67±0.08 | 65.8±0.2 | 64.5±0.2 |
| 091312 | Sep 9 | 22h00m56.1s | -2.8 | 0.39 ± 0.15 | 115.2 | 88.1 | 47.41±0.09 | 39.57±0.08 | 65.8±0.2 | 64.6±0.2 |
| 091313 | Sep 9 | 22h01m17.6s | -5.7 | 7.5 ± 3.0 | 118.2 | 87.3 | 47.32±0.08 | 39.37±0.08 | 65.7±0.2 | 64.4±0.2 |
| 091314 | Sep 9 | 22h02m23.9s | -0.5 | 0.03 ± 0.01 | 105.1 | 95.9 | 47.51±0.08 | 39.76±0.08 | 65.8±0.2 | 64.5±0.2 |
| 091315 | Sep 9 | 22h04m01.2s | -2.2 | 0.20 ± 0.08 | 112.9 | 97.2 | 47.60±0.09 | 39.67±0.08 | 65.9±0.2 | 64.6±0.2 |
| 091316 | Sep 9 | 22h04m20.3s | -5.2 | 4.58 ± 1.86 | 116.7 | 95.2 | 47.90±0.09 | 39.71±0.09 | 66.0±0.2 | 64.7±0.2 |
| 091317 | Sep 9 | 22h04m46.8s | 1.5 | 0.01 ± 0.01 | 100.5 | 96.3 | 47.80±0.10 | 39.82±0.07 | 65.9±0.2 | 64.6±0.2 |
| 091318 | Sep 9 | 22h05m04.0s | -0.6 | 0.04 ± 0.02 | 115.6 | 107.1 | 47.81±0.09 | 39.47±0.08 | 65.9±0.2 | 64.6±0.2 |
| 091319 | Sep 9 | 22h08m10.7s | -3.0 | 0.48 ± 0.20 | 111.9 | 95.1 | 47.40±0.11 | 39.78±0.07 | 65.7±0.2 | 64.5±0.2 |
| 091320 | Sep 9 | 22h09m57.0s | -1.8 | 0.14 ± 0.05 | 113.0 | 97.0 | 47.88±0.10 | 39.48±0.08 | 65.9±0.2 | 64.6±0.2 |
| 091321 | Sep 9 | 22h13m24.3s | -4.6 | 2.5 ± 1.0 | 115.2 | 86.1 | 47.38±0.09 | 39.79±0.07 | 65.7±0.2 | 64.4±0.2 |
| 091322 | Sep 9 | 22h14m05.1s | -0.1 | 0.03 ± 0.01 | 113.2 | 88.4 | 48.00±0.09 | 39.49±0.09 | 66.0±0.2 | 64.7±0.2 |
| 091323 | Sep 9 | 22h16m00.4s | 0.0 | 0.01 ± 0.01 | 101.3 | 96.1 | 47.84±0.09 | 39.52±0.08 | 65.9±0.2 | 64.6±0.2 |
| 091324 | Sep 9 | 22h16m48.9s | 0.5 | 0.01 ± 0.01 | 101.7 | 97.4 | 47.66±0.09 | 39.89±0.07 | 65.8±0.2 | 64.6±0.2 |
| 091325 | Sep 9 | 22h17m19.4s | -2.1 | 0.20 ± 0.08 | 115.1 | 99.1 | 47.86±0.10 | 39.41±0.09 | 65.9±0.2 | 64.6±0.2 |
| 091326 | Sep 9 | 22h28m32.1s | -5.1 | 4.14 ± 1.68 | 120.0 | 95.1 | 47.73±0.10 | 39.78±0.09 | 65.8±0.2 | 64.5±0.2 |
| 091327 | Sep 9 | 22h34m10.6s | -5.8 | 7.76 ± 3.14 | 118.6 | 93.4 | 47.95±0.10 | 39.65±0.09 | 65.8±0.2 | 64.6±0.2 |
| 091328 | Sep 9 | 22h49m01.2s | -4.9 | 3.28 ± 1.33 | 114.8 | 92.2 | 47.61±0.09 | 39.45±0.09 | 65.9±0.2 | 64.7±0.2 |
| 091329 | Sep 9 | 22h52m36.7s | -4.8 | 2.87 ± 1.16 | 114.9 | 90.4 | 48.05±0.07 | 39.34±0.07 | 66.0±0.2 | 64.8±0.2 |
| 091330 | Sep 9 | 23h01m55.9s | 1.0 | 0.01 ± 0.01 | 103.6 | 98.7 | 47.72±0.09 | 39.66±0.07 | 65.8±0.2 | 64.6±0.2 |
| 091331 | Sep 9 | 23h17m15.1s | -5.3 | 5.23 ± 2.12 | 117.8 | 90.7 | 47.77±0.09 | 39.68±0.07 | 65.8±0.2 | 64.6±0.2 |
| 091332 | Sep 9 | 23h23m59.8s | -4.0 | 1.42 ± 0.58 | 109.7 | 95.6 | 47.79±0.09 | 39.59±0.07 | 65.9±0.2 | 64.7±0.2 |
| 091333 | Sep 9 | 23h53m01.0s | -3.3 | 0.64 ± 0.26 | 112.6 | 94.7 | 48.05±0.08 | 39.70±0.07 | 66.0±0.2 | 64.8±0.2 |
| 101301 | Sep 10 | 1h46m40.3s | -4.7 | 2.77 ± 1.12 | 109.4 | 85.8 | 47.82±0.12 | 39.11±0.09 | 65.9±0.2 | 64.8±0.2 |
| 101302 | Sep 10 | 2h45m22.9s | 1.8 | 0.01 ± 0.01 | 105.2 | 97.2 | 47.74±0.13 | 39.63±0.10 | 65.7±0.2 | 64.6±0.2 |
| 101303 | Sep 10 | 3h16m23.7s | -6.1 | 11.6 ± 4.7 | 120.7 | 84.4 | 47.98±0.15 | 39.58±0.12 | 65.8±0.2 | 64.8±0.2 |
| 101304 | Sep 10 | 4h07m48.6s | -4.1 | 1.57 ± 0.64 | 113.9 | 92.1 | 48.09±0.10 | 39.74±0.10 | 65.6±0.2 | 64.6±0.2 |
| 101305 | Sep 10 | 4h32m30.8s | -6.4 | 16.1 ± 6.5 | 115.9 | 85.6 | 47.54±0.10 | 39.67±0.10 | 65.5±0.2 | 64.6±0.2 |
| 101306 | Sep 10 | 5h16m42.5s | -5.6 | 6.73 ± 2.77 | 113.8 | 89.6 | 47.78±0.10 | 39.68±0.10 | 65.6±0.2 | 64.7±0.2 |
| 111301 | Sep 11 | 3h43m32.7s | -3.9 | 1.25 ± 0.50 | 116.2 | 95.6 | 48.12±0.11 | 39.75±0.09 | 65.4±0.2 | 64.4±0.2 |
| 111302 | Sep 11 | 4h52m52.0s | -5.2 | 4.73 ± 1.92 | 115.8 | 90.1 | 49.27±0.10 | 39.66±0.09 | 65.6±0.2 | 64.7±0.2 |
| 111303 | Sep 11 | 5h06m46.6s | -4.3 | 0.86 ± 0.75 | 114.5 | 91.7 | 48.44±0.12 | 40.13±0.10 | 65.3±0.2 | 64.4±0.2 |
| 121301 | Sep 12 | 0h26m38.1s | -2.0 | 0.20 ± 0.08 | 108.5 | 97.8 | 49.37±0.11 | 40.77±0.10 | 65.6±0.2 | 64.4±0.2 |
| 121302 | Sep 12 | 0h37m57.9s | -4.9 | 3.50 ± 1.42 | 105.7 | 89.3 | 49.15±0.10 | 40.10±0.12 | 65.7±0.2 | 64.5±0.2 |
| 121303 | Sep 12 | 2h11m22.1s | -1.8 | 0.16 ± 0.06 | 105.8 | 90.6 | 49.41±0.10 | 40.74±0.10 | 65.4±0.2 | 64.3±0.2 |
| 121304 | Sep 12 | 23h24m55.5s | -5.9 | 9.11 ± 3.69 | 119.5 | 90.7 | 50.00±0.08 | 41.10±0.09 | 65.4±0.2 | 64.2±0.2 |

Table 3. Orbital data (J2000) for the September ε-Perseid meteors listed in Table 2, averaged orbit for N = 60 SPE meteors and averaged orbit for meteors recorded during the outburst (N=28).

| Meteor code | a (AU) | e | i (º) | Ω $\pm 10^{-5}$ (º) | ω (º) | q (AU) | $T_J$ |
|---|---|---|---|---|---|---|---|
| 011301 | 28.4±14.5 | 0.972±0.014 | 140.29±0.11 | 158.71763 | 236.1±0.1 | 0.789±0.002 | -0.65±0.37 |
| 011302 | 35.9±23.5 | 0.978±0.014 | 142.71±0.10 | 159.38597 | 237.0±0.5 | 0.781±0.002 | -0.72±0.47 |
| 011303 | 29.9±16.2 | 0.974±0.014 | 141.55±0.10 | 159.48774 | 238.1±0.5 | 0.774±0.003 | -0.67±0.39 |
| 021301 | 37.3±25.3 | 0.979±0.014 | 140.59±0.12 | 159.56395 | 236.6±0.5 | 0.784±0.002 | -0.70±0.47 |
| 021302 | 19.2±6.6 | 0.959±0.013 | 141.13±0.14 | 159.62377 | 238.2±0.6 | 0.775±0.003 | -0.57±0.27 |
| 031301 | 30.2±16.7 | 0.975±0.013 | 142.20±0.18 | 160.53871 | 243.4±0.6 | 0.733±0.003 | -0.66±0.39 |
| 031302 | 26.6±13.0 | 0.971±0.014 | 140.55±0.20 | 160.63800 | 240.4±0.6 | 0.757±0.003 | -0.63±0.35 |
| 041301 | 24.8±11.3 | 0.970±0.014 | 142.70±0.20 | 161.49337 | 242.7±0.6 | 0.739±0.003 | -0.63±0.34 |
| 051301 | 52.7±51.4 | 0.985±0.013 | 141.75±0.20 | 162.47092 | 241.3±0.6 | 0.747±0.004 | -0.74±0.65 |
| 051302 | 27.3±13.7 | 0.973±0.013 | 140.42±0.20 | 163.27478 | 242.9±0.6 | 0.737±0.004 | -0.62±0.36 |
| 061301 | 29.9±16.2 | 0.975±0.014 | 140.69±0.18 | 163.59031 | 243.5±0.6 | 0.732±0.004 | -0.64±0.38 |
| 081301 | 35.9±23.2 | 0.980±0.013 | 139.53±0.16 | 165.42414 | 245.6±0.6 | 0.714±0.003 | -0.65±0.43 |
| 081302 | 28.2±14.5 | 0.974±0.013 | 140.89±0.18 | 165.46988 | 244.2±0.6 | 0.726±0.004 | -0.63±0.36 |
| 081303 | 34.3±21.7 | 0.979±0.014 | 140.73±0.23 | 165.49119 | 245.4±0.6 | 0.716±0.004 | -0.65±0.43 |
| 091301 | 38.3±26.7 | 0.981±0.013 | 140.24±0.16 | 166.30917 | 243.7±0.6 | 0.729±0.003 | -0.67±0.47 |
| 091302 | 35.4±22.6 | 0.980±0.013 | 140.09±0.16 | 166.33540 | 244.1±0.6 | 0.726±0.003 | -0.66±0.43 |
| 091303 | 56.6±58.2 | 0.987±0.013 | 140.27±0.17 | 166.36164 | 243.9±0.6 | 0.726±0.003 | -0.72±0.66 |
| 091304 | 31.5±17.9 | 0.977±0.014 | 140.04±0.16 | 166.36597 | 244.4±0.6 | 0.724±0.003 | -0.64±0.39 |
| 091305 | 34.6±21.7 | 0.978±0.013 | 140.49±0.17 | 166.40383 | 243.7±0.6 | 0.730±0.003 | -0.66±0.43 |
| 091306 | 40.4±29.7 | 0.982±0.013 | 139.74±0.18 | 167.17936 | 244.7±0.6 | 0.721±0.003 | -0.67±0.48 |
| 091307 | 34.6±21.7 | 0.979±0.013 | 139.30±0.19 | 167.18599 | 244.5±0.5 | 0.723±0.003 | -0.64±0.43 |
| 091308 | 52.3±48.1 | 0.986±0.012 | 140.00±0.14 | 167.18811 | 244.6±0.5 | 0.721±0.003 | -0.70±0.60 |
| 091309 | 33.8±20.7 | 0.978±0.013 | 139.06±0.14 | 167.18831 | 243.7±0.6 | 0.728±0.003 | -0.64±0.41 |
| 091310 | 41.5±31.4 | 0.982±0.013 | 138.92±0.17 | 167.18923 | 244.0±0.6 | 0.726±0.003 | -0.67±0.49 |
| 091311 | 34.3±21.3 | 0.980±0.013 | 139.10±0.16 | 167.19093 | 244.8±0.6 | 0.721±0.003 | -0.64±0.41 |
| 091312 | 35.7±23.1 | 0.979±0.012 | 138.95±0.17 | 167.19106 | 245.1±0.6 | 0.718±0.003 | -0.64±0.43 |
| 091313 | 26.7±12.8 | 0.973±0.013 | 139.13±0.14 | 167.19130 | 246.0±0.6 | 0.712±0.003 | -0.60±0.33 |
| 091314 | 36.7±24.4 | 0.980±0.013 | 138.71±0.12 | 167.19204 | 244.5±0.6 | 0.722±0.003 | -0.64±0.44 |
| 091315 | 44.7±36.5 | 0.983±0.013 | 139.00±0.16 | 167.19314 | 244.4±0.6 | 0.723±0.003 | -0.67±0.52 |
| 091316 | 44.9±36.8 | 0.983±0.013 | 139.21±0.13 | 167.19336 | 243.6±0.6 | 0.729±0.003 | -0.68±0.54 |
| 091317 | 39.9±29.0 | 0.981±0.013 | 138.88±0.14 | 167.19365 | 243.8±0.6 | 0.728±0.003 | -0.66±0.48 |
| 091318 | 32.4±19.0 | 0.977±0.013 | 139.41±0.17 | 167.19386 | 244.5±0.6 | 0.723±0.003 | -0.63±0.40 |
| 091319 | 30.5±16.9 | 0.976±0.013 | 138.56±0.15 | 167.19594 | 245.0±0.6 | 0.720±0.003 | -0.61±0.37 |
| 091320 | 30.4±16.8 | 0.976±0.013 | 139.45±0.18 | 167.19716 | 244.4±0.6 | 0.724±0.004 | -0.62±0.38 |
| 091321 | 31.1±17.5 | 0.977±0.013 | 138.53±0.17 | 167.19947 | 245.0±0.6 | 0.720±0.003 | -0.61±0.38 |
| 091322 | 37.6±25.8 | 0.981±0.013 | 139.56±0.12 | 167.19995 | 244.0±0.5 | 0.726±0.003 | -0.66±0.45 |
| 091323 | 32.2±18.7 | 0.978±0.013 | 139.36±0.14 | 167.20124 | 244.4±0.6 | 0.724±0.003 | -0.63±0.39 |
| 091324 | 34.4±21.5 | 0.979±0.013 | 138.63±0.14 | 167.20177 | 244.1±0.6 | 0.726±0.003 | -0.64±0.42 |
| 091325 | 30.0±16.3 | 0.976±0.013 | 139.54±0.16 | 167.20213 | 244.6±0.6 | 0.722±0.003 | -0.62±0.37 |
| 091326 | 31.4±17.9 | 0.977±0.013 | 138.80±0.17 | 167.20969 | 244.2±0.6 | 0.725±0.003 | -0.62±0.39 |
| 091327 | 24.9±11.2 | 0.971±0.013 | 139.19±0.17 | 167.21349 | 244.3±0.6 | 0.726±0.003 | -0.58±0.32 |
| 091328 | 39.9±29.1 | 0.982±0.013 | 139.29±0.18 | 167.22352 | 244.9±0.6 | 0.719±0.003 | -0.66±0.48 |
| 091329 | 34.1±21.0 | 0.979±0.013 | 139.81±0.12 | 167.22594 | 244.3±0.5 | 0.725±0.003 | -0.65±0.42 |
| 091330 | 30.4±16.8 | 0.976±0.013 | 139.00±0.15 | 167.23222 | 244.6±0.6 | 0.723±0.003 | -0.62±0.38 |
| 091331 | 30.1±16.5 | 0.976±0.013 | 139.01±0.15 | 167.24256 | 244.5±0.6 | 0.724±0.003 | -0.62±0.37 |
| 091332 | 38.2±26.5 | 0.981±0.013 | 139.21±0.16 | 167.24711 | 244.4±0.6 | 0.723±0.003 | -0.66±0.46 |
| 091333 | 49.8±44.9 | 0.985±0.013 | 139.24±0.14 | 167.26670 | 243.5±0.5 | 0.730±0.003 | -0.70±0.58 |
| 101301 | 45.4±37.7 | 0.984±0.014 | 139.95±0.18 | 167.34341 | 245.2±0.6 | 0.716±0.003 | -0.68±0.54 |
| 101302 | 40.9±30.6 | 0.982±0.014 | 139.03±0.20 | 167.38301 | 244.7±0.6 | 0.720±0.004 | -0.66±0.49 |
| 101303 | 58.1±62.0 | 0.987±0.014 | 139.25±0.33 | 167.40395 | 244.2±0.6 | 0.724±0.004 | -0.70±0.67 |
| 101304 | 34.7±21.8 | 0.979±0.013 | 139.10±0.19 | 167.43864 | 244.1±0.6 | 0.725±0.003 | -0.64±0.42 |
| 101305 | 48.1±41.0 | 0.985±0.013 | 138.80±0.20 | 167.45531 | 245.2±0.6 | 0.716±0.004 | -0.67±0.55 |
| 101306 | 54.6±54.0 | 0.986±0.014 | 139.00±0.21 | 167.45531 | 244.6±0.6 | 0.721±0.004 | -0.70±0.62 |
| 111301 | 38.1±26.3 | 0.981±0.012 | 138.60±0.20 | 168.39366 | 246.7±0.6 | 0.704±0.004 | -0.64±0.45 |
| 111302 | 31.3±17.7 | 0.977±0.013 | 139.65±0.19 | 168.44117 | 244.8±0.6 | 0.721±0.003 | -0.63±0.39 |
| 111303 | 37.1±25.0 | 0.981±0.013 | 138.23±0.21 | 168.45053 | 245.6±0.6 | 0.714±0.004 | -0.64±0.44 |
| 121301 | 48.2±42.4 | 0.985±0.013 | 137.57±0.19 | 169.23389 | 244.7±0.6 | 0.720±0.003 | -0.66±0.54 |
| 121302 | 29.0±15.1 | 0.975±0.012 | 138.82±0.18 | 168.54458 | 244.8±0.6 | 0.721±0.003 | -0.61±0.36 |
| 121303 | 33.5±20.3 | 0.978±0.013 | 137.56±0.19 | 169.30464 | 245.1±0.6 | 0.718±0.004 | -0.62±0.40 |
| 121304 | 35.7±23.0 | 0.980±0.013 | 137.02±0.16 | 170.16520 | 245.8±0.6 | 0.712±0.003 | -0.62±0.41 |
| Average (N=60) | 36.2±25.1 | 0.978±0.013 | 139.60±0.16 | 166.10643 | 243.7±0.6 | 0.729±0.003 | -0.65±0.44 |
| Outburst (N=28) | 35.8±24.0 | 0.979±0.013 | 139.15±0.16 | 167.20462 | 244.4±0.6 | 0.723±0.003 | -0.64±0.43 |

## FIGURES

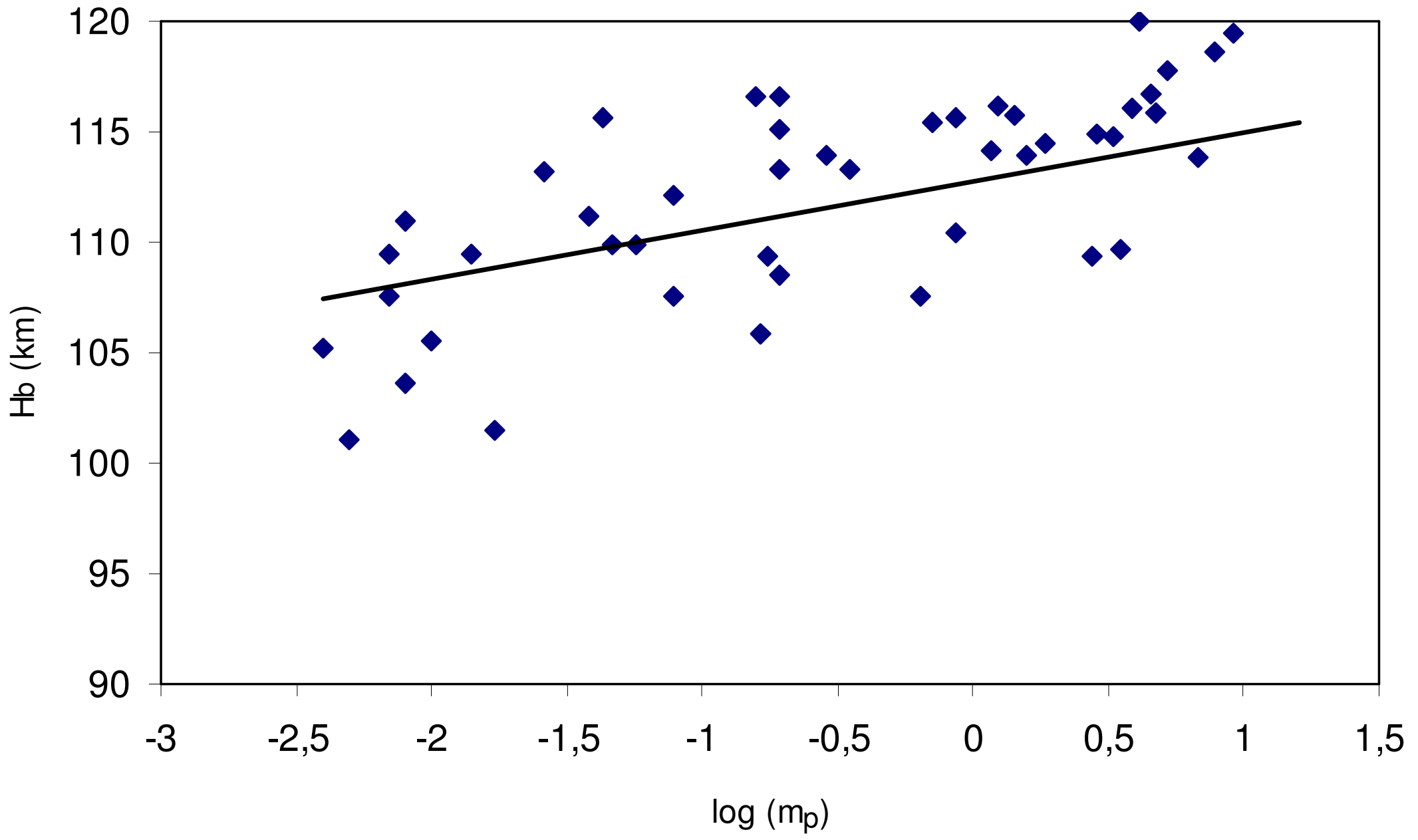

Figure 1. Meteor beginning height $H_b$ vs. logarithm of the photometric mass $m_p$ of the meteoroid. Solid line: linear fit for measured data.

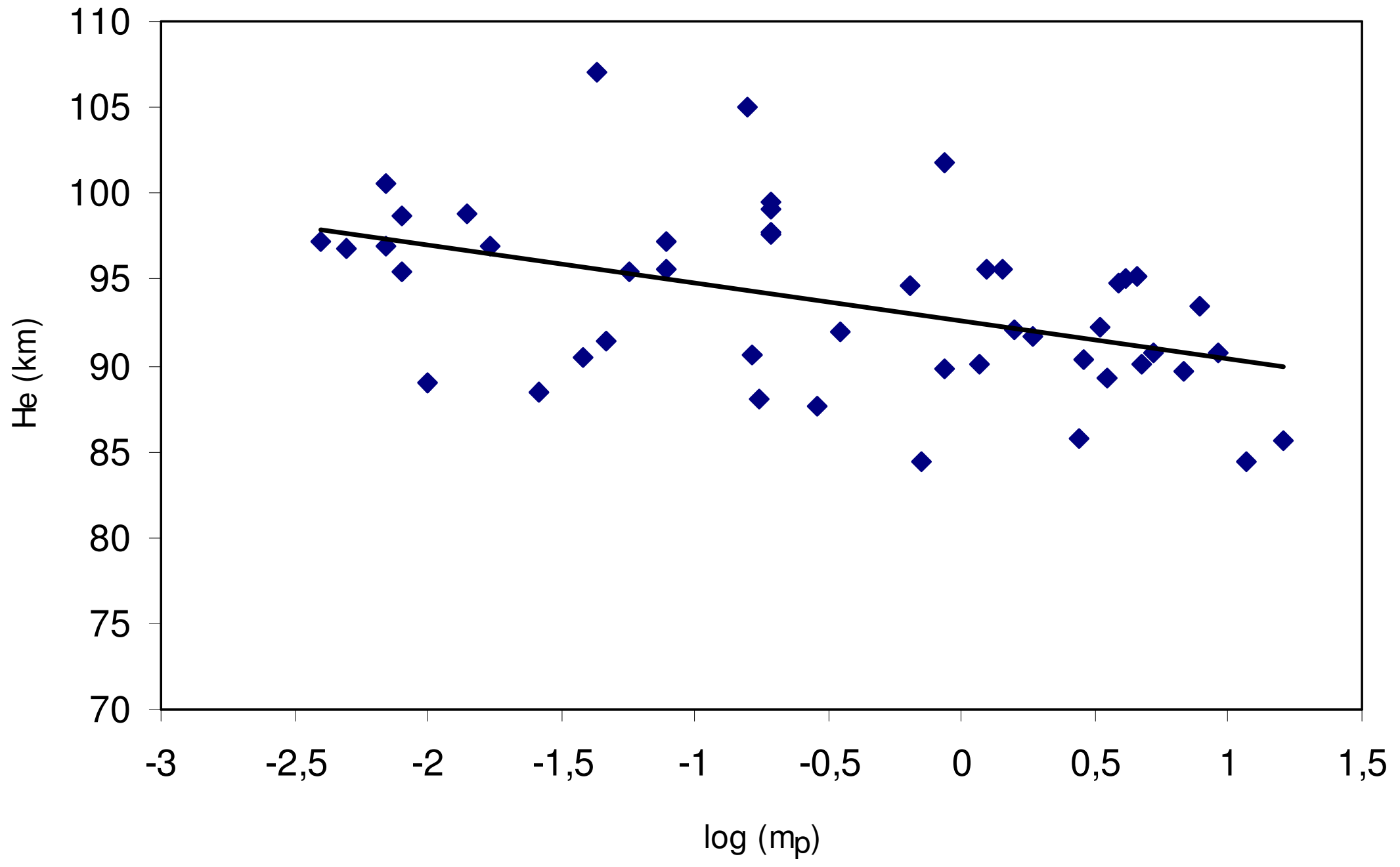

Figure 2. Meteor beginning height $H_b$ vs. logarithm of the photometric mass $m_p$ of the meteoroid. Solid line: linear fit for measured data.

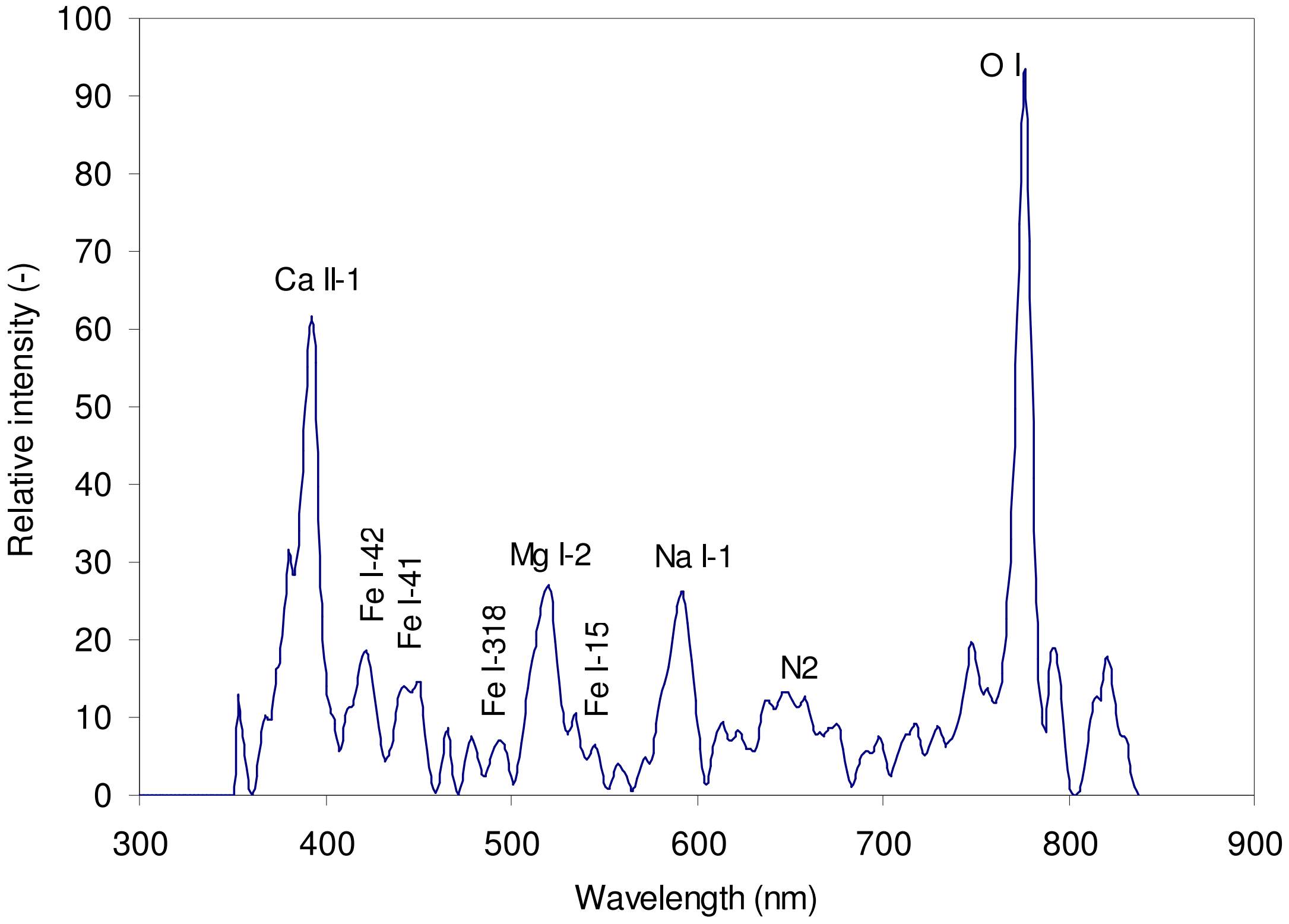

Figure 3. Calibrated emission spectrum of the SPE091327 meteor.

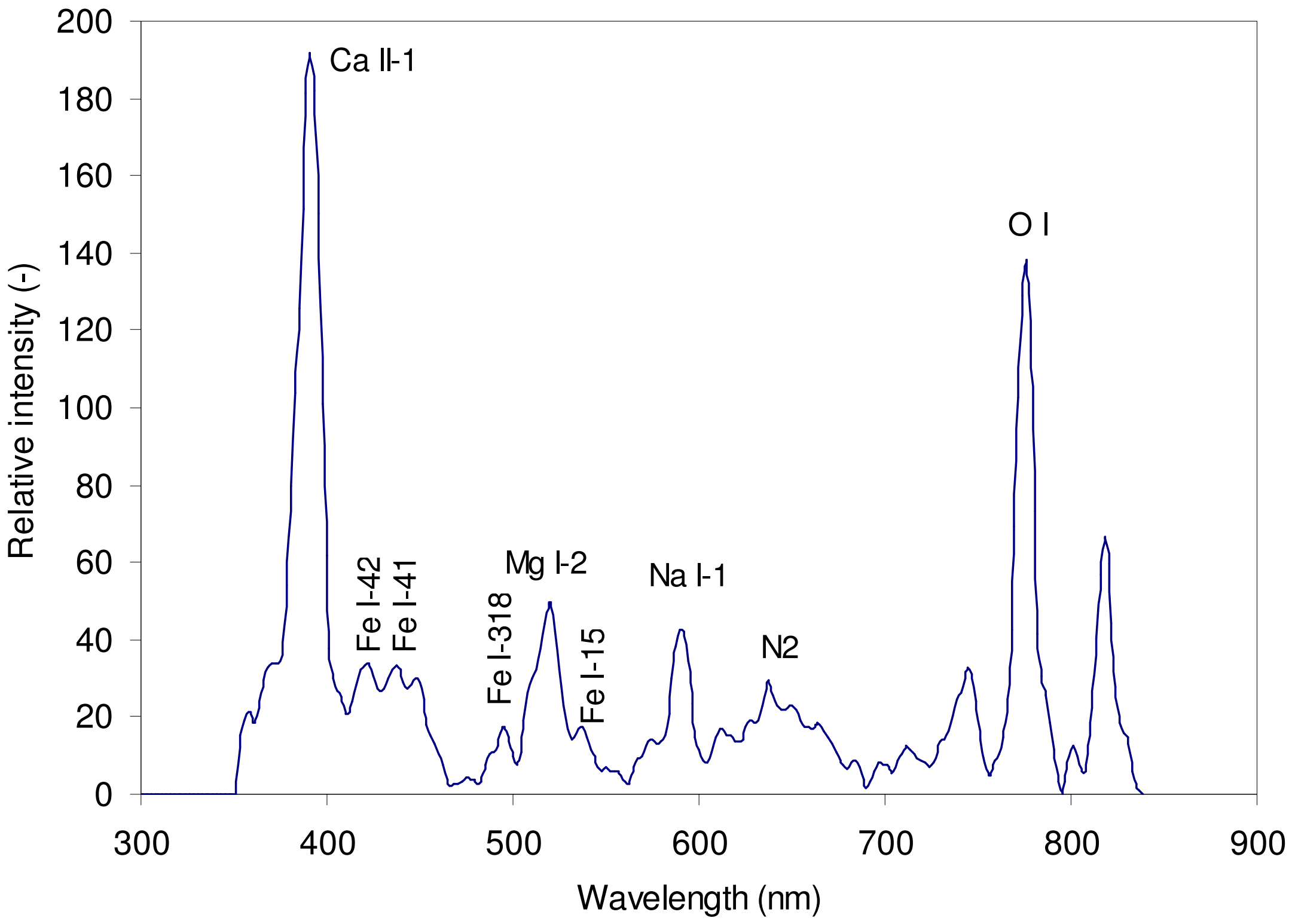

Figure 4. Calibrated emission spectrum of the SPE091331 meteor.

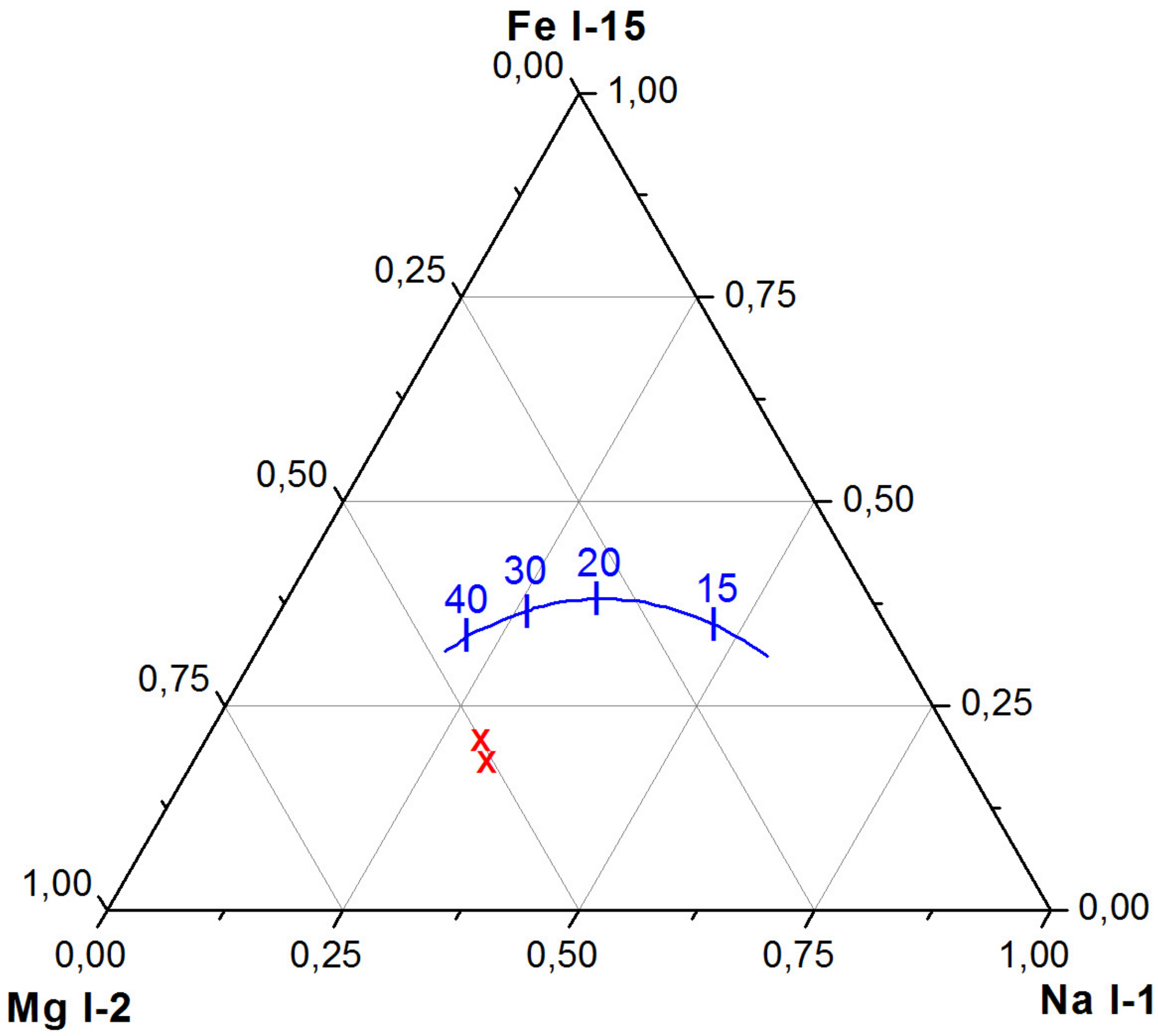

Figure 5. Expected relative intensity (solid line), as a function of meteor velocity (in km $s^{-1}$), of the Na I-1, Mg I-2 and Fe I-15 multiplets for chondritic meteoroids (Borovička et al., 2005). Crosses: experimental relative intensities obtained for the SPE091327 and SPE091331 spectra.